# The dominance of non-electron-phonon charge carrier interaction in highly-compressed superhydrides

Evgeny F. Talantsev[1,2]

[1]M.N. Mikheev Institute of Metal Physics, Ural Branch, Russian Academy of Sciences, 18, S. Kovalevskoy St., Ekaterinburg, 620108, Russia

[2]NANOTECH Centre, Ural Federal University, 19 Mira St., Ekaterinburg, 620002, Russia

**Abstract**

The primary mechanism governing the emergence of near-room-temperature superconductivity in superhydrides is widely accepted to be the electron-phonon interaction. If so, the temperature dependent resistance, $R(T)$, in these materials should obey the Bloch-Grüneisen equation, where the power-law exponent, $p$, should be equal to the exact integer value of $p=5$. On the other hand, there is a well-established theoretical result that pure electron-magnon interaction should be manifested by $p=3$, and $p=2$ is the value for pure electron-electron interaction. Here we aimed to reveal the type of charge carrier interaction in the layered transition metal dichalcogenides $PdTe_2$, high-entropy alloy $(ScZrNb)_{0.65}[RhPd]_{0.35}$ and highly-compressed elemental boron and superhydrides $H_3S$, $LaH_x$, $PrH_9$ and $BaH_{12}$ by fitting the temperature dependent resistance of these materials to the Bloch–Grüneisen equation, where the power-law exponent, $p$, is a free-fitting parameter. In the result, we showed that the high-entropy alloy $(ScZrNb)_{0.65}[RhPd]_{0.35}$ exhibited pure electron-phonon mediated superconductivity with $p = 4.9 \pm 0.4$. Unexpectedly we revealed that all studied superhydrides exhibit $1.8 < p < 3.2$. This implies that it is unlikely that the electron-phonon interaction is the primary mechanism for the Cooper pairs formation in highly-compressed superhydrides and alternative pairing mechanisms, for instance, the electron-magnon, the electron-polaron, the electron-electron or other, should be considered as the origin for the emergence of near-room-temperature superconductivity in these compounds.

**The dominance of non-electron-phonon charge carrier interaction in highly-compressed superhydrides**

**I. Introduction**

The discovery of the first superhidride/superdeuteride superconductor $Th_4H_{15}$/$Th_4D_{15}$ by Satterthwaite and Toepke [1] was based on a very clear idea, which they expressed as [1]: "…There has been theoretical speculation [2] that metallic hydrogen might be a high-temperature superconductor, in part because of the very high Debye frequency of the proton lattice. With high concentrations of hydrogen in the metal hydrides one would expect lattice modes of high frequency and if there exists an attractive pairing interaction one might expect to find high-temperature superconductivity in these systems also."

Surprisingly enough Satterthwaite and Toepke [1] found that isotopic counterparts $Th_4H_{15}$ and $Th_4D_{15}$ have the same superconducting transition temperature, $T_c$, which is in direct contradiction with the theory of electron-phonon mediated superconductivity [3,4], including the case of hydrogen-rich metallic alloys [5]. Later it was found that the PdH-PdD-PdT system exhibits the inverse isotope effect [6,7], where heavier counterparts have higher superconducting transition temperature in comparison with isotopically lighter compounds.

In this regard, we should mention that, despite a milestone experimental discovery reported by Drozdov *et al* [8] on the near-room-temperature superconductivity (NRTS) in highly-compressed sulphur hydride, recent report by Minkov *et al* [9] showed that the isotope effect in $H_3S$-$D_3S$ system is not as prominent as it was initially reported [8] in 2015. A primary open question here is what the exact stoichiometry of the hydrogen isotopes in NRTS samples is. There is some experimental evidence [10] that hydrogen- and deuterium-based NRTS phases have different hydrogen isotopes stoichiometry and, thus, it is possible that the observed difference in $T_c$ for isotopic counterparts is originated from the stoichiometry, rather than from the fundamental difference in the phonon spectra.

Because the charge carriers can form the Cooper pairs not only by the attractive electron-phonon interaction [11-14], there is a necessity to reaffirm or disprove the electron-phonon mediated superconductivity in the NRTS superhydrides. In an attempt to do this, here we analysed the temperature dependent resistance, $R(T)$, in several highly-compressed superhydrides by using the generalized Bloch-Grüneisen (BG) equation [15,16], where the power-law exponent, $p$, is assumed to be a free-fitting parameter.

In the result we showed the dominance of non-electron-phonon charge carrier interactions (in particular, the electron-magnon and the electron-electron interactions) in highly-compressed $H_3S$, $LaH_x$, $PrH_9$, $CeH_9$, and $BaH_{12}$ superhydrides. This result is in a good accord with reports [17,18] where it was shown that NRTS superconductors are in the unconventional superconductors region of the Uemura plot [19,20]. This implies that superhydride superconductors exhibit a non-electron-phonon pairing mechanism.

## 2. Model description

To determine the type of the charge carrier interaction in NRTS here we proposed to use an advance version of the Bloch-Grüneisen (BG) equation [15,16] which can be represented in the most general form as [21]:

$$R(T) = R_0 + A_1 \cdot T + \sum_n^{2,3,5} A_p \cdot \left(\frac{T}{T_\theta}\right)^p \cdot \int_0^{\frac{T_\theta}{T}} \frac{x^p}{(e^x-1)\cdot(1-e^{-x})} \cdot dx \qquad (1)$$

where $R_0$ is the resistance at $T \rightarrow 0$ K, $T_\theta$ is the Debye temperature, $A_p$ is weighting parameters, and $p$ is the power-law exponent which has theoretical integer value for the following interaction mechanisms [21]:

$$p = \begin{cases} 2 \text{ for the electron} - \text{electron interaction} \\ 3 \text{ for the electron} - \text{magnon interaction} \\ 5 \text{ for the electron} - \text{phonon interaction} \end{cases} \qquad (2)$$

However, as it is pointed in Ref. 22, Eq. 1 has a linear limit for $p \rightarrow 1$:

$$\lim_{p \to 1} \left(\frac{T}{T_\theta}\right)^p \cdot \int_0^{\frac{T_\theta}{T}} \frac{x^p}{(e^x-1)\cdot(1-e^{-x})} \cdot dx \to \left(\frac{B}{T_\theta}\right) \cdot T \tag{3}$$

where $B$ is a constant. Thus, the linear term in Eq. 1 can be represented as an integral part at $p \to 1$ with weighting factor $A_1$:

$$R(T) = R_0 + \sum_p^{(\lim_{p \to 1}),2,3,5} A_p \cdot \left(\frac{T}{T_\theta}\right)^p \cdot \int_0^{\frac{T_\theta}{T}} \frac{x^p}{(e^x-1)\cdot(1-e^{-x})} \cdot dx \tag{4}$$

We should stress that Eq. 4 was never applied for the analysis of experimental $R(T)$ data, because there is a sum of integrals and, thus, the fitting procedure is over-parametrized. However, to the best of the author's knowledge, in all published work (except, the work by Jiang *et al* [23] and another recent report [22]) Eq. 4 was used for the electron-phonon integrand, i.e. $p = 5$ (see, for instance, [24-26]).

One possible way to use Eq. 4 is to reduce the number of integrals to one, but in this integral the power-law exponent, $p$, will be a free-fitting parameter:

$$R(T) = R_0 + A_p \cdot \left(\frac{T}{T_\omega}\right)^p \cdot \int_0^{\frac{T_\omega}{T}} \frac{x^p}{(e^x-1)\cdot(1-e^{-x})} \cdot dx \tag{5}$$

where we introduce the designation of $T_\omega$ for the characteristic temperature for the case when $p$ is a free-fitting parameter, while the designation of $T_\theta$ is kept for $p = 5$.

Thus, the dominant charge carrier interaction mechanism in given materials can be determined from comparison of the deduced free-fitting parameter $p$ with theoretical values for pure cases (Eq. 2).

From the best author's knowledge, Eq. 5 was first used by Jiang *et al* [23] to reveal the dominant interaction mechanism in $Sr_2Cr_3As_2O_2$ ferrimagnet. Resulting mechanism was established as the electron-magnon (with deduced $p = 3.34$ [23]), which was the first confirmation of the approach validity. More recently, in Ref. 22, the approach was applied for pure copper, iron and cobalt (for which $R(T)$ was reported by Matula [27], Teixeira [28], and White and Woods [29]). Also, the approach confirmed the electron-phonon mediated superconductivity in $ReBe_{22}$ (for which $R(T)$ was reported by Shang *et al* [24]). On the other hand, for the highly-compressed superconducting ε-phase of iron (for which $R(T)$ was reported by Shimizu *et al* [30] and by Jaccard *et al* [31]), and for twisted bilayer graphene superlattice (for which $R(T)$ was reported by Polshin *et al* [32]) the non-electron-phonon mechanism of superconductivity was revealed.

Here to fit $R(T)$ data we used the recently proposed equation [33]:

$$R(T) = R_0 + \theta(T_c^{onset} - T) \cdot \left( \frac{R_{norm}}{\left(I_0\left(F \cdot \left(1 - \frac{T}{T_c^{onset}}\right)^{3/2}\right)\right)^2} \right) + \theta(T - T_c^{onset}) \cdot \Bigg( R_{norm} + A \cdot$$

$$\left( \left(\frac{T}{T_\omega}\right)^p \cdot \int_0^{\frac{T_\omega}{T}} \frac{x^p}{(e^x - 1)\cdot(1 - e^{-x})} \cdot dx - \left(\frac{T_c^{onset}}{T_\omega}\right)^p \cdot \int_0^{\frac{T_\omega}{T_c^{onset}}} \frac{x^p}{(e^x - 1)\cdot(1 - e^{-x})} \cdot dx \right) \Bigg) \qquad (6)$$

where instead of fixed $p = 5$ (as it was the case in Ref. 33), here we used $p$ as a free-fitting parameter, $T_c^{onset}$ is free-fitting parameter of the onset of superconducting transition, $R_{norm}$ is the sample resistance at the onset of the transition, $\theta(x)$ is the Heaviside step function, $I_0(x)$ is the zero-order modified Bessel function of the first kind and $F$ is a free-fitting dimensionless parameter. The designation of $T_\theta$ (i.e. the Debye temperature) is kept for the fits with fixed $p = 5$. More details about Eq. 6 (for the case of fixed $p = 5$) can be found elsewhere [21,27,29,34].

It should be noted that there is an alternative approach to analyse $R(T)$ datasets [21,27,29,31,35] by replacing Eq. 5 by simple power-law fitting function:

$$R(T) = R_0 + A_N \cdot T^N \qquad (7)$$

where $N$ is a free-fitting parameter. This approach assumes that the deduced $N$-value (in Eq. 7) is accurate approximation for $p$-value in Eq. 5. Thus, in this assumption, the fitting equation for superconductors is:

$$R(T,B) = R_0 + \theta(T_c^{onset} - T) \cdot \left( \frac{R_{norm}}{\left(I_0\left(F \cdot \left(1 - \frac{T}{T_c^{onset}}\right)^{3/2}\right)\right)^2} \right) + \theta(T - T_c^{onset}) \cdot$$

$$\left(R_{norm} + A_N \cdot (T^N - (T_c^{onset})^N)\right) \tag{8}$$

There is a need to point out a fundamental problem associated with the use of Eqs. 7,8 which is the unit of free-fitting parameter $A_N$. From the logic it should have the unit of Siemens $^{N-1}$, and if (as we show below) $N = 2.69$, the unit should be Siemens $^{1.69}$ or $\frac{1}{Ohm^{1.69}}$. However, there is no physical parameter or measurable value which has the unit of Siemens $^{1.69}$ or $\frac{1}{Ohm^{1.69}}$. This problem can be resolved if Eq. 7 transforms to:

$$R(T) = R_0 + A_N \cdot \left(\frac{T}{T_{char}}\right)^N \tag{9}$$

where $T_{char}$ is some characteristic temperature. However, Eq. 9 cannot be used to fit $R(T)$ data, because parameters $A_N$ (which has unit of Ohm in Eq. 9), $T_{char}$, and $N$ have mutual dependence $\equiv 1$. This can be resolved if $A_N$ in Eq. 9 will be fixed to some conventional value. However, there is no clarity what this value can be and what can be physical meanings for $A_N$ and $T_{char}$ (Eq. 9).

However, to demonstrate that widely used power-law approximant (Eqs. 7,8) cannot be reliable substitution for Eqs. 5,6, we fitted $R(T)$ data for elemental copper and silver, PdTe$_2$, high-entropy alloy (ScZrNb)$_{0.65}$[RhPd]$_{0.35}$ and several highly-compressed superconductors to to Eqs. 5-8 and reported results herein.

## 3. Results

### 3.1. Elemental copper

First, we made a comparison of fits to Eq. 5 and Eq. 7 for temperature dependent resistivity, $\rho(T)$, for pure copper. Experimental $\rho(T)$ data was measured by Teixeira [28] and this dataset was referred as dataset #269 by Matula [29]. For the analysis we used $\rho(T)$ data in two temperature ranges, $1.2\,K \leq T \leq 34.2\,K$ and $1.2\,K \leq T \leq 70.6\,K$, and, thus, both analysed $\rho(T)$ datasets were within low-$T$ ranges of $\frac{T}{T_\theta} \leq \frac{1}{10}$ and of $\frac{T}{T_\theta} \leq \frac{1}{5}$ respectively, where $T_\theta = 320 - 342\,K$ [27,36].

$\rho(T)$ data fits to Eqs. 7,5 for the temperature range of $\frac{T}{T_\theta} \leq \frac{1}{10}$ are shown in Figs. 1(a) and 1(b), where deduced $N = 4.72 \pm 0.02$ and $p = 5.08 \pm 0.03$. These values are in a proximity to expected value of 5. However, $N$-value deduced for $\rho(T)$ data measured at twice wider temperature range ($\frac{T}{T_\theta} \leq \frac{1}{5}$) is significantly different, $N = 3.39 \pm 0.04$ (Fig. 1(c)), while deduced $p = 4.82 \pm 0.02$ remains practically the same (Fig. 1(d)). This result is in agreement with general understanding [21,27-29] that the power-law function (Eq. 7) is a good approximation for Bloch-Grüneisen (BG) equation (Eq. 5) at $\frac{T}{T_\theta} \leq \frac{1}{10}$.

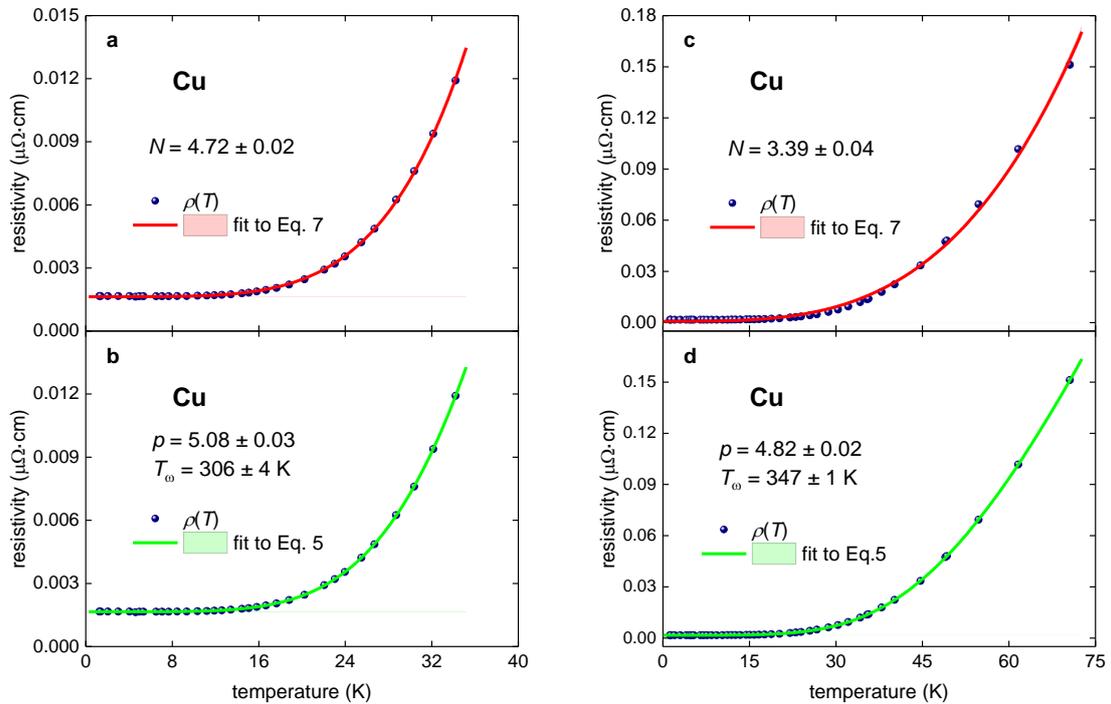

**Figure 1.** $\rho(T)$ data for pure elemental copper reported by Teixeira [28] (this $\rho(T)$ dataset referred as dataset #269 by Matula [27]). **a,b** for temperature range $\frac{T}{T_\theta} \leq \frac{1}{10}$ and **c,d** for temperature range $\frac{T}{T_\theta} \leq \frac{1}{5}$. (a) fit to Eq.7, deduced $N = 4.72 \pm 0.02$, goodness of fit is 0.9998; (b) fit to Eq. 5; deduced $p = 5.08 \pm 0.03$, $T_\omega = 306 \pm 4\,K$, goodness of fit is 1.0. (c) fit to Eq.7, deduced $N = 3.39 \pm 0.04$,

goodness of fit is 0.997; (d) fit to Eq. 5; deduced $p = 4.82 \pm 0.02$, $T_\omega = 347 \pm 1\ K$, goodness of fit is 1.0. 95% confidence bands are shown by shaded areas.

Deduced Debye temperatures for two temperature ranges, i.e. $T_\theta(T < 34\ K) = 306 \pm 4\ K$ and of $T_\theta(T < 70\ K) = 347 \pm 1\ K$, are in a reasonable agreement with reported $T_\theta = 320 - 342\ K$ for pure copper [36].

### 3.2. Elemental silver

To demonstrate that the approach described in previous Section is reliable, in this Section we performed the analysis for pure elemental silver. In Fig. 2 we showed experimental $\rho(T)$ data reported by Teixeira [28] (this dataset was referred as dataset 103 by Matula [27]) and data fits to Eq. 5,7. For the analysis we used $\rho(T)$ data in two temperature ranges of $1.28\ K \leq T \leq 22.8\ K$ (which is low-$T$ region, because $\frac{T}{T_\theta} \leq \frac{1}{10}$ (where $T_\theta = 221 - 228\ K$ [27,36])), and of $1.28\ K \leq T \leq 113\ K$ (which is within medium-$T$ range of $\frac{T}{T_\theta} \leq \frac{1}{2}$).

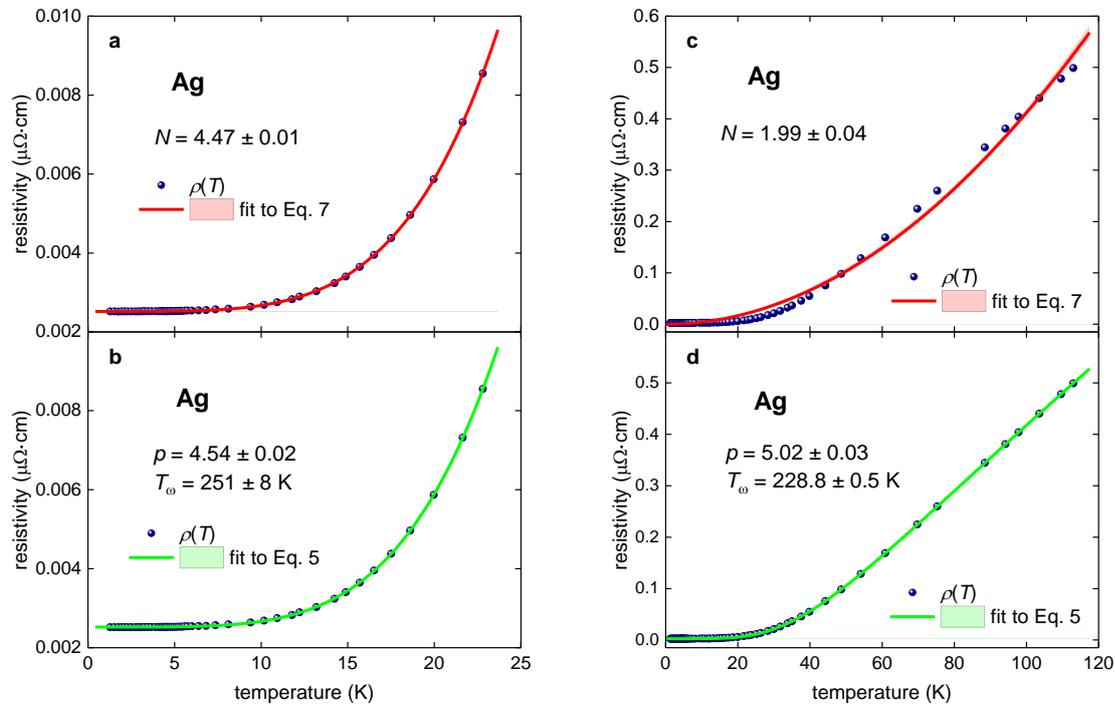

**Figure 2.** $\rho(T)$ data for pure elemental copper reported by Teixeira [28] (this $\rho(T)$ dataset referred as dataset #103 by Matula [27]). **a,b** for temperature range $\frac{T}{T_\theta} \leq \frac{1}{10}$ and **c,d** for temperature range $\frac{T}{T_\theta} \leq$

$\frac{1}{2}$. (a) fit to Eq.7, deduced $N = 4.47 \pm 0.01$, goodness of fit is 0.99997; (b) fit to Eq. 5, deduced $p = 4.54 \pm 0.02$, $T_\omega = 251 \pm 8\ K$, goodness of fit is 1.0. (c) fit to Eq.7, deduced $N = 1.99 \pm 0.04$, goodness of fit is 0.994; (d) fit to Eq. 5; deduced $p = 5.02 \pm 0.03$, $T_\omega = 228.8 \pm 0.5\ K$, goodness of fit is 1.0. 95% confidence bands are shown by shaded areas.

It can be seen (Fig. 2(c)) that the fit to Eq. 7 has low quality and free-fitting parameter $N = 1.99 \pm 0.04$ is very different with expected value of 5, while the fit to Eq. 5 (Fig. 2(d)) has an excellent quality and $p = 5.02 \pm 0.03$. In addition, deduced Debye temperature, $T_\theta = 228.8 \pm 0.5\ K$ (Fig. 2(d)), is in excellent agreement with reported $T_\theta = 221 - 228\ K$, which was deduced from the specific heat measurements by Gschneider [36].

### 3.3. PdTe$_2$

Layered transition metal dichalcogenides (TMDCs) possess various unusual electronic properties, including the emerge of the superconductivity [37,38] from the inherent instabilities of the stripe charge ordered phase [38,39]. Generalized Bloch-Grüneisen equation (Eq. 5) has been never applied for the analysis of temperature dependent resistivity in TMDCs and here we performed the analysis for PdTe$_2$. In Fig. 3 we show $R(T)/R(300K)$ data for single crystal PdTe$_2$ at ambient pressure recently reported by Yang *et al* [40]. These authors claimed that their analysis showed unusual $N = 4$ value for the fit to the power-law function of $R(T)/R(300K)$ data. Our fits to power-law function (Eq. 7, showed in Figs. 3(a,d)) of the same data set revealed that *N*-value is much lower ($1.57 \leq N \leq 2.35$) and it has remarkably strong dependence from the $R(T)/R(300K)$ temperature range, chosen for analysis. Also, the fit does not have high quality.

To demonstrate that generalized Bloch-Grüneisen equation (Eq. 5) can be a better research tool to analyse temperature dependent resistivity data in Figs. 3(b,e) we show the data fits to Eq. 5, where *p* was fixed to $p = 5$ and in Figs. 3(c,f) the data fits to Eq. 5, where *p* was free-fitting parameter. Deduced $T_\theta = 153 - 164\ K$ and $T_\omega = 191 - 205\ K$ is varying within 7% despite a fact that the temperature range for analysed $R(T)/R(300K)$ dataset is varying by factor of two. Free-fitting parameter $p = 3.08 - 3.30$ (Figs. 3(c,f)) is also varying in a narrow range and it is definitely well below $N = 4$ reported by Yang *et al* [40].

It should be noted that there are two primary requirements for the $R(T)$ dataset to be reliably fitted to Eq. 5. The first requirement is that $R(T)$ data should be measured at as wide as possible range of $\frac{T}{T_\omega}$ ratio. Truly, it can be seen in Figs. 1(b,d), 2(b,d) and 3(c,f) that the uncertainties for deduced $T_\omega$ values are dramatically narrower for $R(T)$ datasets taken at wider temperature range. The origin for this reduction is that Eq. 5 has the term of:

$$\left(\frac{T}{T_\omega}\right)^p \cdot \int_0^{\frac{T_\omega}{T}} \frac{x^p}{(e^x-1)\cdot(1-e^{-x})} \cdot dx \qquad (10)$$

and it is obvious that if $R(T)$ data covers wider $\frac{T}{T_\omega}$ range, than the full term (i.e., Eq. 10) will be more accurately calculated. Also, when p value is free-fitting parameter it has some uncertainty and, thus, full integral (Eq. 10), where the second free-fitting parameter is $T_\omega$, can be calculated with wider uncertainty.

Also, there is the second requirement that $R(T)$ dataset should contain many datapoints to calculate the same integral (Eq. 10) with better accuracy. Because, both these requirements are not really satisfied for NRTS superconductors, for which $T_\omega > 1,500\ K$ and $R(T)$ datasets are conventionally measuring up to $T = 295$ K only, detailed discussion of this problem for NRTS materials is given below in Section 3.6.

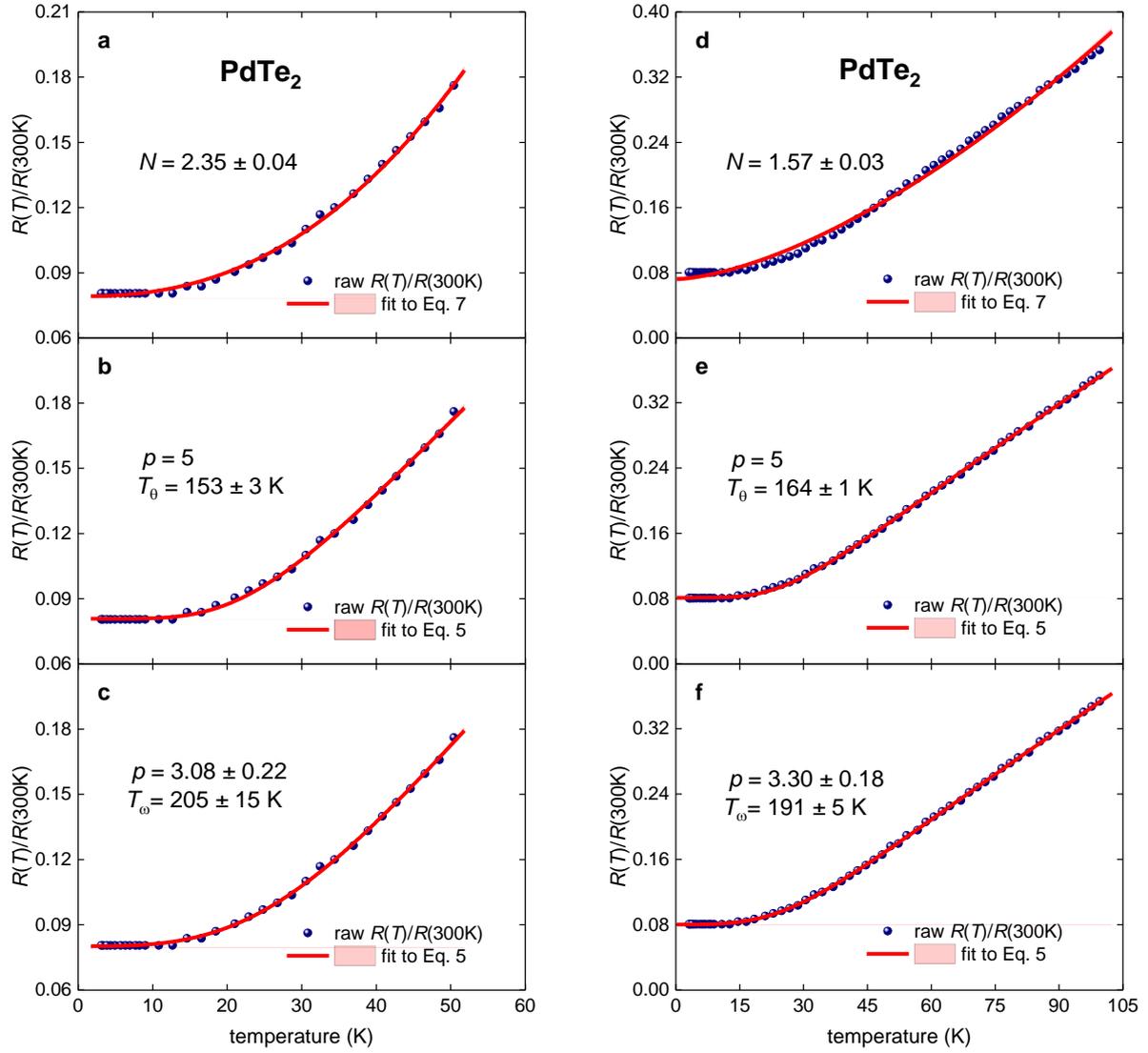

**Figure 3.** $R(T)/R(300K)$ data for TMDC PdTe$_2$ at ambient pressure reported by Yang *et al* [40]. **a-c** - temperature range $\frac{T}{T_\theta} \leq \frac{1}{3}$ and **d-f** - temperature range $\frac{T}{T_\theta} \leq \frac{2}{3}$. (a) fit to Eq.7, deduced $N = 2.35 \pm 0.04$, goodness of fit is 0.9982; (b) fit to Eq. 5, $p \equiv 5$, $T_\theta = 153 \pm 3\,K$, goodness of fit is 0.9987. (c) fit to Eq.5, deduced $p = 3.08 \pm 0.22$, $T_\omega = 205 \pm 15\,K$, goodness of fit is 0.9992; (d) fit to Eq.7, deduced $N = 1.57 \pm 0.03$, goodness of fit is 0.9964; (e) fit to Eq. 5, $p \equiv 5$, $T_\theta = 164 \pm 1\,K$, goodness of fit is 0.9998. (f) fit to Eq.5, deduced $p = 3.30 \pm 0.18$, $T_\omega = 191 \pm 5\,K$, goodness of fit is 0.99986. 95% confidence bands are shown by shaded areas.

### 3.4. High-entropy alloy (ScZrNb)$_{0.65}$[RhPd]$_{0.35}$

Before Eqs. 5,6 will apply to hydrogen-rich superconductors, there is a necessity to show that these equations are applicable for low-temperature superconductors. In Fig. 4 we applied Eqs. 6,8 to fit $R(T)$ data for high-entropy alloy (ScZrNb)$_{0.65}$[RhPd]$_{0.35}$ for which raw experimental $R(T)$ data was reported by Stolze *et al* [41]. It should be mentioned that recently we fitted this $R(T)$ dataset to Eq. 5 for fixed $p = 5$ [33], and here we make the $p$ parameter to be free.

It can be seen (Fig. 4) that both fitting curves have high quality. However, the fit to power-law model (Eq. 8, Fig. 4(a)) revealed $N = 1.2$, which is remarkably different from fit to BG model (Eq. 6, Fig. 4(b)) for which the deduced $p = 4.9 \pm 0.4$. This implies that the superconducting state in high-entropy alloy (ScZrNb)$_{0.65}$[RhPd]$_{0.35}$ emerges solely from the electron-phonon interaction. Also, from results presented in Fig. 4 it can be seen, that the power-law function (Eq. 8) cannot be accepted to be valid approximation for integrated form of Bloch-Grüneisen equation (Eqs. 5,6).

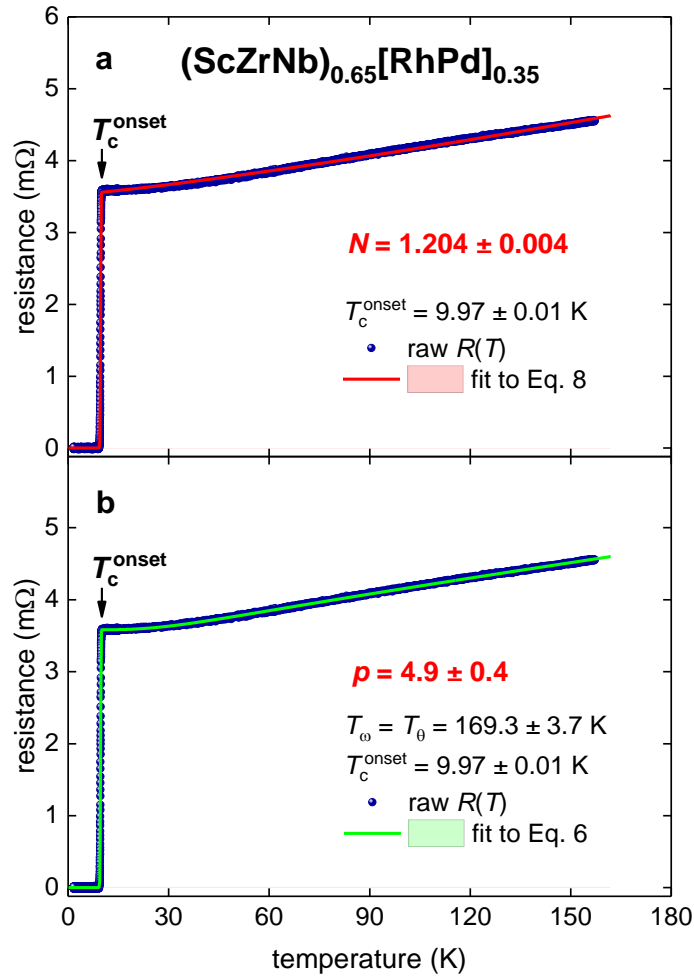

**Figure 4.** $R(T)$ data for high-entropy alloy (ScZrNb)$_{0.65}$[RhPd]$_{0.35}$ reported by Stolze *et al* [41] and data fit to Eqs. 6,8. (a) fit to Eq. 8, $N = 1.204 \pm 0.004$, goodness of fit is 0.99991; (b) $p = 4.9 \pm 0.4$, $T_\omega = 169.3 \pm 3.7\ K$, goodness of fit is 0.99997. 95% confidence bands are narrower than the fitting curve.

### 3.5. α-Ga phase of highly-compressed boron

Now we turn to highly-compressed superconductors and in Fig. 4 we showed $R(T)$ data for elemental boron pressurized at $P = 240$ GPa (reported by Eremets *et al* [42]). Oganov *et al* [43] showed that elemental boron at pressure above $P = 89$ GPa exists in so-called α-Ga phase and, thus, the superconducting state emerges in this boron phase.

$R(T)$ fit to Eq. 8 is shown in Fig. 5(a), for which deduced $N = 1.32 \pm 0.02$. It should be stressed that the fit to Eq. 6 can be performed also for fixed $p = 5$ value (Fig. 5,b). This fit has a high quality (the goodness of fit is 0.9977) and deduced $T_\theta = 318 \pm 7\ K$ is in the expected range.

However, when *p* was chosen to be a free-fitting parameter, the deduced value $p = 2.3 \pm 0.3$ indicates that the electron-electron interaction is the dominant in this superconductor. To explain this unexpected result, one can take into account that elemental boron is situated between metals and insulators in the Mendeleev's table. Despite this element exhibits three valence electrons, which, in principle, can lead to metallic conductivity, the electrons are sufficiently localized and at ambient conditions boron is an insulator [43]. Thus, it is not a real surprise, that the electron-electron interaction dominates in this material even at high-pressure.

There is a need to discuss two issues. The first one is that the power-law approximation (Eq. 8) of the BG integral form (Eq. 6) cannot be considered as a valid substitution because deduced $N = 1.32 \pm 0.02$ (Eq. 8) is significantly different from $p = 2.3 \pm 0.3$ (Eq. 6). The second issue is that deduced characteristic temperature, $T_\omega = 435 \pm 38\ K$, has much wider uncertainty range in comparison with $T_\theta = 318 \pm 7\ K$. As we already mentioned it above, this is because $T_\omega > T_\theta$ and, thus, $R(T)$ dataset is analysed at effectively narrower reduced temperature range (which has the upper limit of $\frac{T}{T_\omega}$) in comparison with the case when the same dataset is analysed at reduced temperature range with the upper limit of $\frac{T}{T_\theta}$. In addition, the uncertainty of free-fitting parameter $p$ causes the increase in the uncertainty of free-fitting parameter $T_\omega$, while at fixed $p$ value the uncertainty in calculating integral in Eq. 10 is solely dependent from the single parameter $T_\theta$ (see for details Section 3.3).

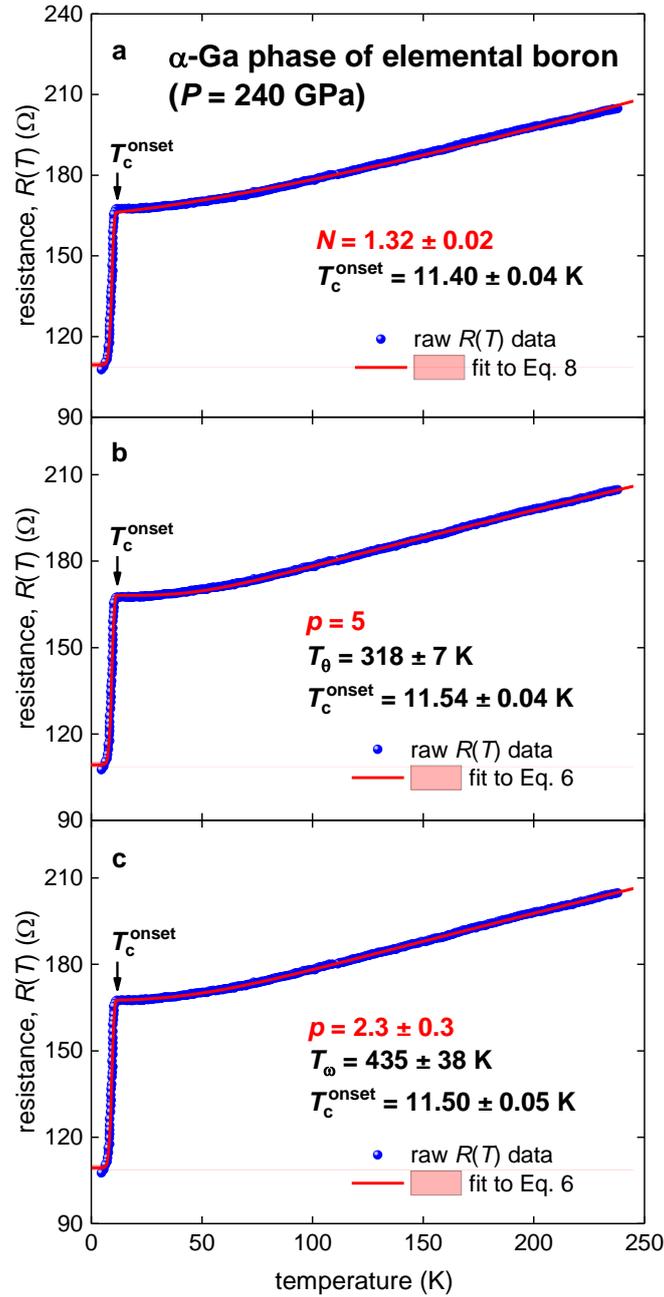

**Figure 5.** Resistance data, $R(T)$, and fits to Eqs. 6,8 for highly-compressed elemental boron (raw data reported by Eremets *et al* [42]). (a) fit to Eq. 8, $N = 1.32 \pm 0.02$, goodness of fit is 0.9972. (b) fit to Eq. 6, $p = 5$ (fixed), $T_\theta = 318 \pm 7\ K$, goodness of fit is 0.9977. (c) fit to Eq. 6, $p = 2.3 \pm 0.3$, $T_\omega = 435 \pm 38\ K$ goodness of fit is 0.9978. 95% confidence bands are shown by pink shaded areas.

## 3.6. *Im-3m*-phase of highly-compressed H₃S

Now we turn for the analysis of $R(T)$ data for NRTS H$_3$S for which extended raw datasets are freely available by Mozaffari *et al* [44]. In their Fig. 1, Mozaffari *et al* [44] reported $R(T)$ curves for *Im-3m*-H$_3$S samples subjected to pressure of $P = 155\ GPa$ and of $P = 160\ GPa$. In Fig. 6 we showed $R(T)$ curves and fits to Eqs. 8 for sample **07/2018** compressed at $P = 155$ GPa which was aged for two (**09/2018**) and four (**11/2018**) consequence months.

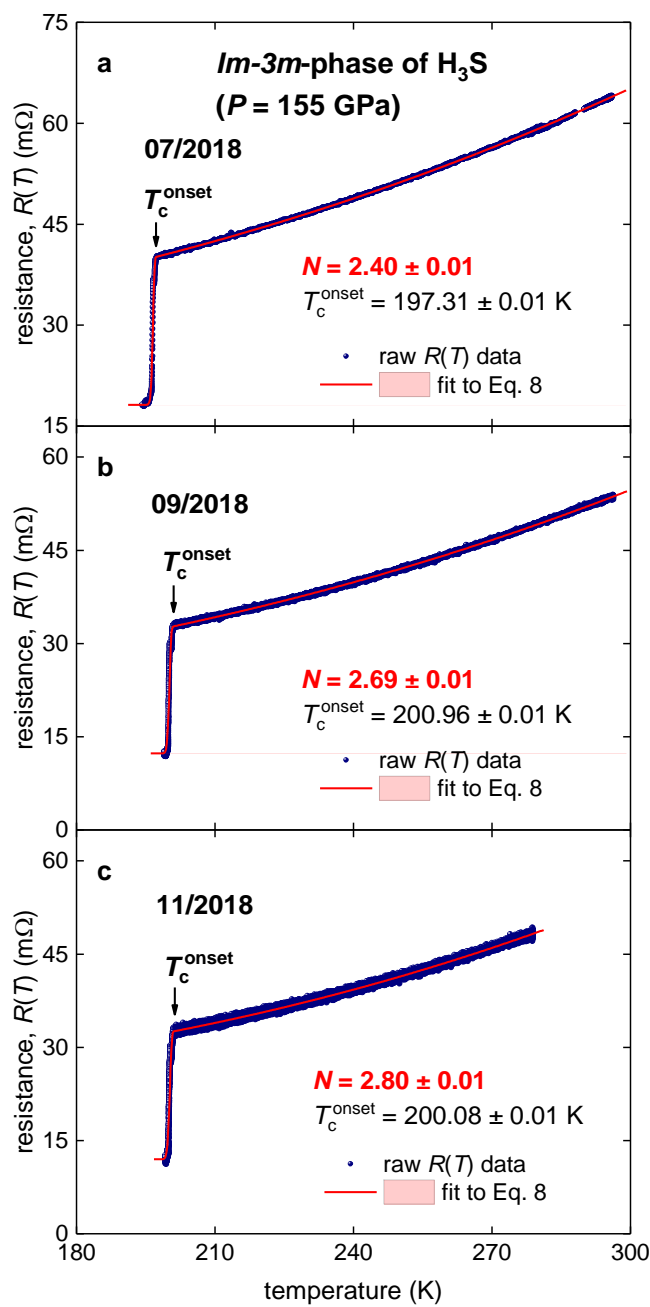

**Figure 6.** Resistance data, $R(T)$, and fits to Eq. 8 for highly-compressed *Im-3m*-H$_3$S phase (raw data reported by Mozaffari *et al* [44]). (a) Sample 07/2018, $N = 2.40 \pm 0.01$, goodness of fit is 0.9995; (b)

Sample 09/2018, $N = 2.69 \pm 0.01$, goodness of fit is 0.9972; (c) Sample 09/2018, $N = 2.80 \pm 0.01$, goodness of fit is 0.9936. 95% confidence bands are shown as shaded areas.

It should be noted that $R(T)$ curves for this sample have multiple-step transition, which are shown in Fig. 7. We fitted the transition for Sample **07/2018** to two-step fitting function in our previous work (Fig. 9 in Ref. 33). Here, to analyse three $R(T)$ datasets, we cut off $R(T)$ curves at $T_{\text{cut off}}$ (Fig. 7) and fitted $R(T > T_{\text{cut off}})$ to Eq. 8 (Fig. 7) and Eq. 6 (Fig. 8).

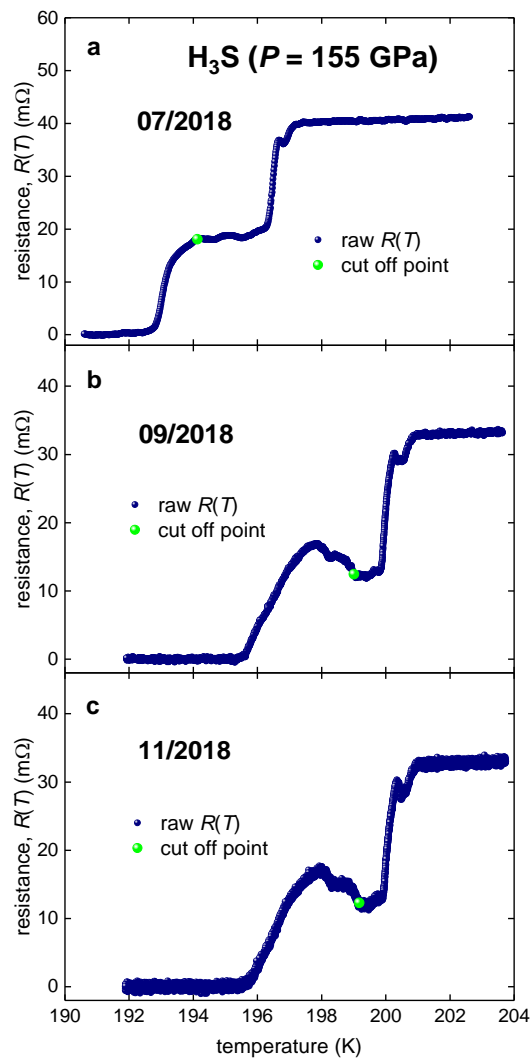

**Figure 7.** Resistive transition curves for $H_3S$ samples 07/2018 (a), 09/2018 (b) and 11/2018 (c) (raw data reported by Mozaffari *et al* [44]). Green balls indicate the cut-off points for the $R(T)$ datasets analysed in Fig. 6.

It should be noted that Mozaffari *et al* [44] fitted $R(T)$ dataset for Sample **07/2018** to the parabolic function:

$$R(T) = R_0 + A \cdot T^2 \qquad (11)$$

and based on a fact that the fit has a good quality, the authors [44] made a conclusion that the charge carrier interaction in $H_3S$ is dominated by strong interaction between electrons and very high energy optical phonon. This mechanism initially was proposed by Capitani *et al* [45].

However, the fit to Eq. 11 cannot prove that $R(T)$ has a parabolic dependence because the power-law exponent is already fixed to $N = 2$. In Fig. 6 we presented fits to Eq. 8 for Samples **07/2018**, **09/2018** and **11/2018**, where $N$ was a free-fitting parameter. Based on results of fits (Fig. 6), one can conclude that there are no experimental evidences that $R(T)$ of $H_3S$ has parabolic temperature dependence.

To further investigate this issue we return to work by Capitani *et al* [45] who fitted $R(T)$ curves for two $H_3S$ samples (Sample A ($P$ = 160 GPa) and Sample B ($P$ = 150 GPa) [45]) to a model based on so-called the elastic residual scattering rate $\gamma_r$. Capitani *et al* [45] deduced $\gamma_r$ = 28 meV and $\gamma_r$ = 135 meV for Sample A and Sample B respectively, which means that main fitting parameter is different by factor of ~5 for these two samples.

In Fig. 8 we presented the fits to Eq. 8 of $R(T)$ datasets for Sample A and Sample B [45]. Because Sample A exhibits two-step transition $R(T)$ dataset was cut off (Fig. 8(a)). From Figs. 7,8 one can conclude that $R(T)$ curves of highly-compressed $H_3S$ cannot be characterized by parabolic temperature dependence.

We showed in previous papers [33,34], that full $R(T)$ transition curves for several NRTS materials can be fitted to BG equation (Eq. 6) where $p = 5$ (fixed). Here we extended this list by $H_3S$ samples of **07/2018**, **09/2018** and **11/2018** [44] (all compressed at $P = 155$ GPa, Fig. 9) and samples **07/2018** ($P = 160$ GPa [44]), Sample A ($P = 160$ GPa [45]) and Sample B ($P = 150$ GPa [44]) (which are shown in Fig. 10). In Table I we collected main deduced parameters for *Im-3m*-H$_3$S samples for which *R*(*T*) data is available to date. It can be seen that the Debye temperature, $T_\theta$, is varying within a narrow range for all H$_3$S samples.

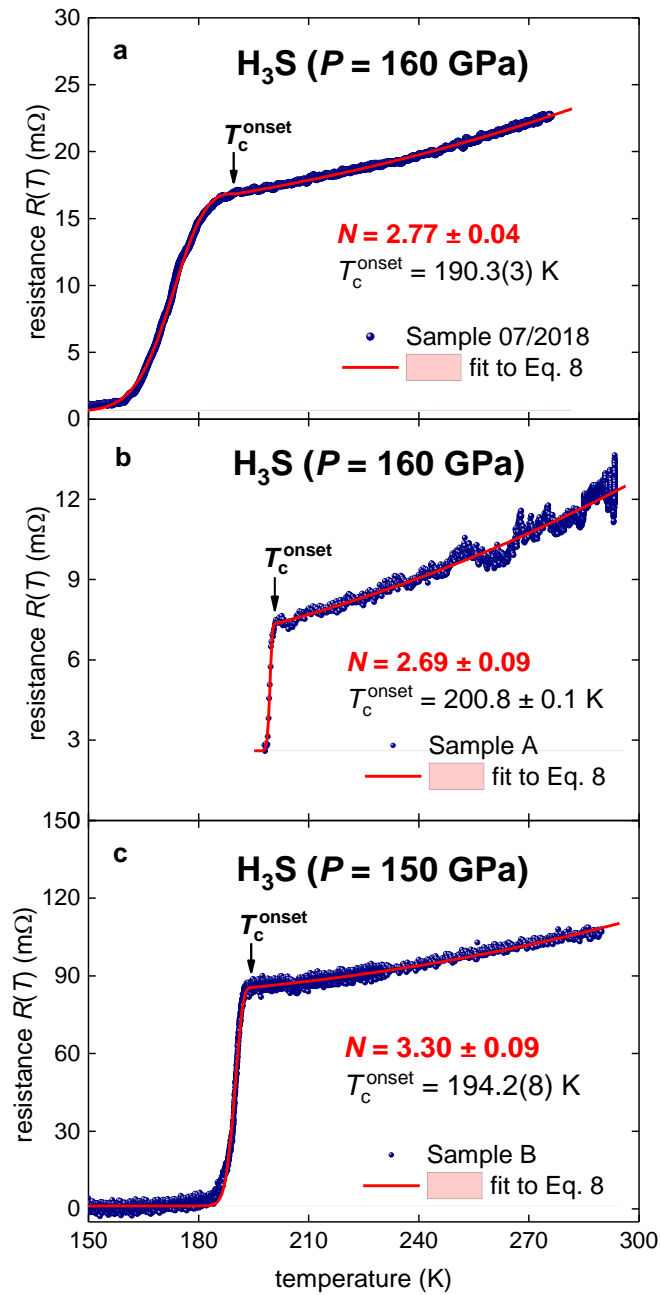

**Figure 8.** Resistance data, *R*(*T*), and fits to Eq. 8 for highly-compressed *Im-3m*-H$_3$S phase. Raw data reported by (a) Mozaffari *et al* [44] and (b,c) Capitani *et al* [45]. (a) Sample 07/2018 (*P* = 160 GPa), *N* = 2.77 ± 0.04, goodness of fit is 0.9397; (b) Sample A (*P* = 160 GPa), *N* = 2.69 ± 0.09, goodness of fit is 0.9337; (c) Sample B (*P* = 150 GPa), *N* = 3.30 ± 0.09, goodness of fit is 0.9987. 95% confidence bands are shown by shaded areas.

**Table I.** Main deduced parameters for *Im-3m*-H$_3$S phase.

| Sample ID | Pressure (GPa) | $T_c^{onset}$/T | N-value (Eq. 8) | $T_\theta$ (K) (Eq. 6, $p$ = 5) | $T_c^{onset}/T_\theta$ | $T_{upper\,exp}/T_\theta$ | $T_\omega$ (K) (Eq. 6, $p$ is free) | Free-fitting parameter $p$ (Eq. 6) |
|---|---|---|---|---|---|---|---|---|
| Sample B [45] | 150 | 194.3 | 3.30±0.09 | 1,764±61 | 0.110 | 0.166 | | |
| 07/2018 [44] | 155 | 197.3 | 2.40±0.01 | 1,427±2 | 0.138 | 0.207 | 2,707±184 | 2.44±0.03 |
| 09/2018 [44] | 155 | 201.0 | 2.69±0.01 | 1,560±5 | 0.129 | 0.188 | | |
| 11/2018 [44] | 155 | 201.1 | 2.80±0.01 | 1,552±6 | 0.130 | 0.180 | | |
| 07/2018 [44] | 160 | 190.3 | 2.77±0.04 | 1,487±19 | 0.128 | 0.186 | | |
| Sample A [45] | 160 | 200.8 | 2.69±0.09 | 1,618±50 | 0.124 | 0.179 | | |

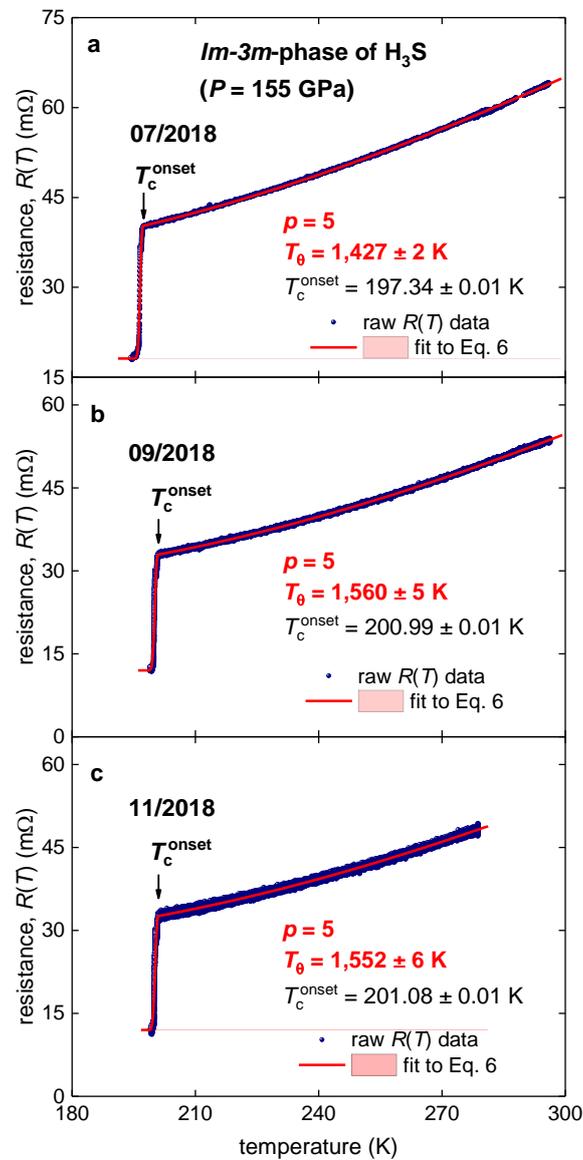

**Figure 9.** Resistance data, *R*(*T*), and fits to Eq. 6 (for $p = 5$ (fixed)) for highly-compressed *Im-3m*-H$_3$S phase (raw data reported by Mozaffari *et al* [44]). (a) Sample 07/2018, $T_\theta = 1427 \pm 2\,K$, goodness of fit is 0.9995; (b) Sample 09/2018, $T_\theta = 1560 \pm 5\,K$, goodness of fit is 0.9972; (c) Sample 09/2018, $T_\theta = 1552 \pm 6\,K$, goodness of fit is 0.9936. 95% confidence bands are shown as shaded areas.

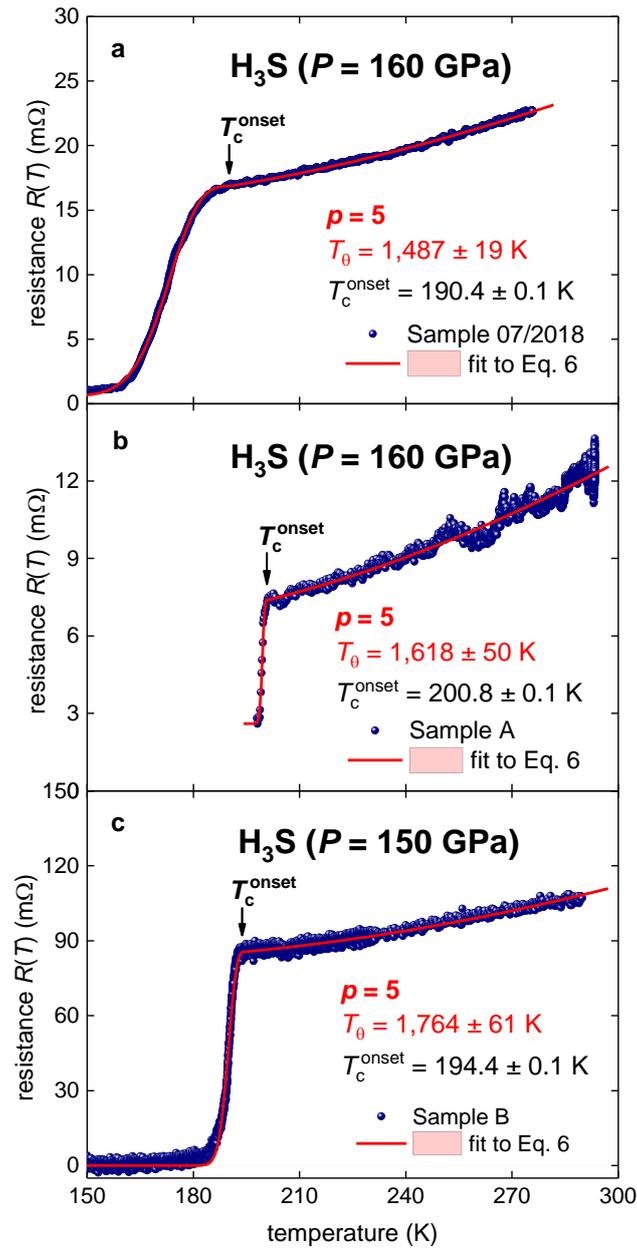

**Figure 10.** Resistance data, $R(T)$, and fits to Eq. 6 (for $p = 5$ (fixed)) for highly-compressed $Im\text{-}3m$-$H_3S$ phase. $R(T)$ data reported by (a) Mozaffari *et al* [44] and (b,c) by Capitani *et al* [45]. (a) Sample 07/2018 ($P = 160$ GPa), $T_\theta = 1487 \pm 19\ K$, goodness of fit is 0.9997; (b) Sample A ($P = 160$ GPa), $T_\theta = 1618 \pm 50\ K$, goodness of fit is 0.9337; (c) Sample B ($P = 150$ GPa), $T_\theta = 1764 \pm 61\ K$, goodness of fit is 0.9984. 95% confidence bands are shown as shaded areas.

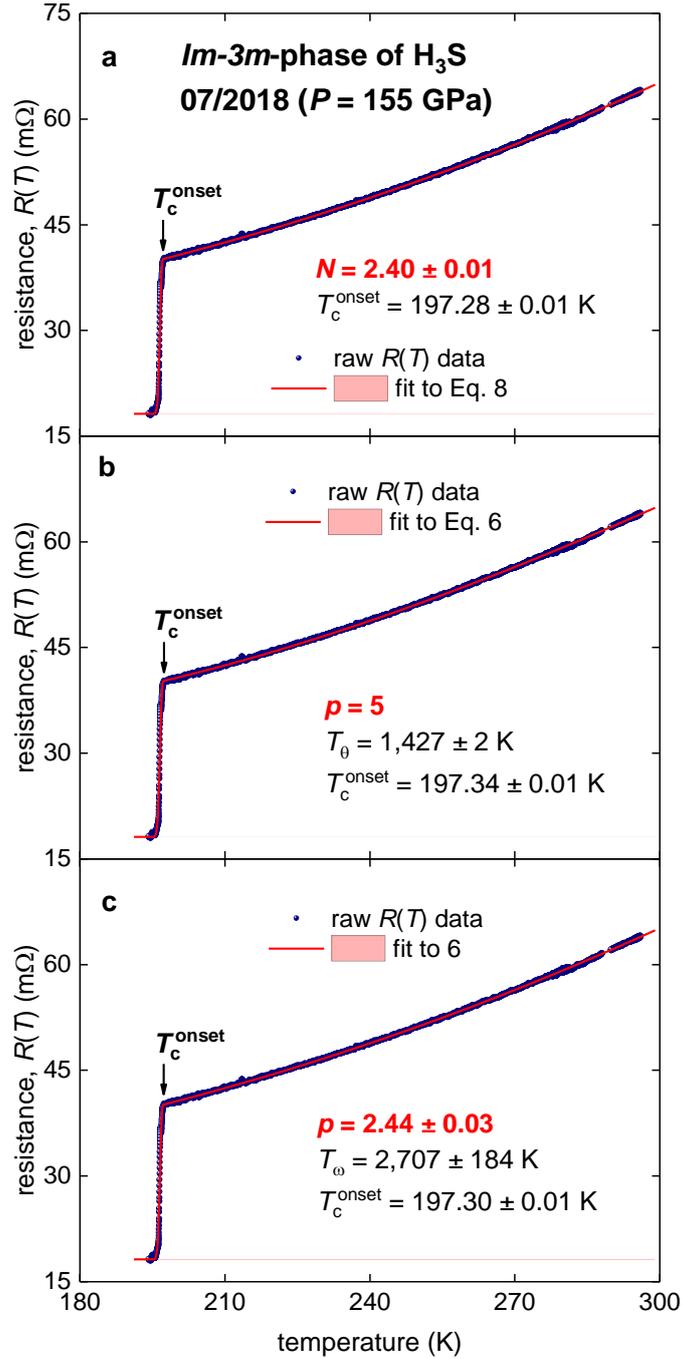

**Figure 11.** Resistance data, $R(T)$, and fits to Eqs. 6,8 for highly-compressed *Im-3m*-$H_3S$ phase (raw data for this sample designated as ***155 GPa: 07/2018*** reported by Mozaffari *et al* [44]). (a) fit to Eq. 8, $N = 2.39 \pm 0.01$, goodness of fit is 0.9996; (b) fit to Eq. 6, $p = 5$ (fixed), $T_\theta = 1427 \pm 2\,K$, goodness of fit is 0.9994. (c) $p = 2.44 \pm 0.03$, $T_\omega = 2707 \pm 184\,K$, goodness of fit is 0.9995. 95% confidence bands are narrower that the fitting curve.

It can be seen in Fig. 10 (b,c) that Sample A and Sample B exhibit very close $T_\theta$ values, which correlate with their close $T_c^{onset}$ values. However, this result is in a contrast with a model proposed by Capitani *et al* [45] who assumed that primary parameter of their model, which is the elastic residual scattering rate γ_r, should be $\gamma_r$ = 28 meV and $\gamma_r$ = 135 meV for Sample A and Sample B respectively. We argue herein that advanced BG model (Eq. 6) is sufficient to deduce the primary parameter of NRTS (which is $T_\theta$ or $T_\omega$) and to fit $R(T)$ curves for these materials [33,34].

It can be seen in Figs.9,10 and Table I) that fits to Eq. 6 at fixed $p$ = 5 exhibit high qualities and deduced $T_\theta$ are in agreement with respectful characteristic temperatures for the phonon spectrum reported by first-principles calculations for this compound [47-53]. However, in the case when $p$ is a free-fitting parameter, its value is reduced to close proximity to the electron-electron interaction value of $p = 2.44 \pm 0.03$.

However, only one of six $R(T)$ datasets (i.e. 07/2018) can be fitted to Eq. 6 (Fig. 11) when $p$ is a free-fitting parameter. Other fits were either diverged, either parameters uncertainties were larger than the deduced values, either parameters dependency reaches value $\equiv 1$.

We already discussed some issues of this problem in Sections 3.2,3.3. In addition to already mentioned issues, we should stress that to be fitted to Eqs. 5,6 the $R(T)$ dataset should be measured in substantially wide temperature range of $T_c^{onset} \leq T \leq T_\omega$. For example, $R(T)$ datasets for considered in Sections 3.1, 3.2 pure Cu and Ag cover $\frac{1}{300} \lesssim \frac{T}{T_\omega} \lesssim \frac{1}{2}$, which allows to converge that fits and to deduce $p$ and $T_\omega$ values with small uncertainties.

For superconductors, the lower $\frac{T}{T_\omega}$ boundary is naturally limited by the normal part of $R(T)$ curve:

$$\frac{T_c^{onset}}{T_\omega} \lesssim \frac{T}{T_\omega} \qquad (12)$$

because all $R(T < T_c^{onset})$ data points are excluded from the data fit to entire BG integral (Eq. 10). This is important property of the model, because if even $R(T)$ curve has multiple-step transition, the deduced free-fitting parameters $p$ and $T_\theta$ are not affected by this feature, due to $p$ and $T_\theta$ deduced for $R(T > T_c^{onset})$.

If for weak-coupling superconductors the lower boundary (Eq. 12) is reasonably small [54-57]:

$$0.005 \lesssim \frac{T_c^{onset}}{T_\theta} \lesssim 0.04, \qquad (13)$$

and cannot create a significant problem to accurately calculate the BG integral in Eq. 6, for strong-coupling superconductors (including, in accordance with the most widely accepted point of view, all NRTS [47-53,58]) this lower limit is much higher and for $H_3S$ this limit is (see Table I):

$$0.11 \leq \frac{T_c^{onset}}{T_\theta}. \qquad (14)$$

Despite a fact that the lower limit (Eq. 12) is the fundamental model limit to analyse $R(T)$ datasets for superconductors, its impact on the accuracy of computations can be significantly reduced if $R(T)$ dataset measured at significant part of full temperature range of $\frac{T}{T_\omega}$, . But, to date, all reported $R(T)$ datasets for NRST (including reported by Mozaffari *et al*. [44], Drozdov *et al* [59]) have the upper limit of $T_{upper\ exp} \sim 295\ K$, which is in five cases of six is too low to allow to calculate the integral part of the Eq. 6 with required accuracy (see Table I). Truly, for sample 07/2018 deduced $T_\omega = 2707 \pm 184\ K$, the highest temperature is $T_{upper\ exp} \sim 295.98\ K$ [44], and thus:

$$\frac{T_{upper\ exp}}{T_\omega} = 0.109. \qquad (15)$$

Because other $H_3S$ samples in Table I have higher $T_\theta$ it is naturally to propose that $T_\omega$ for those samples will be also higher, and thus the ratio of $\frac{T_{upper\ exp}}{T_\omega}$ will be even lower that for sample 07/2018 (Eq. 15). It should be mentioned that the lower boundary for 07/2018 samples is:

$$0.073 = \frac{T_c}{T_\omega}. \qquad (16)$$

Thus, $R(T)$ for sample 07/2018 (the only dataset, from six available, for which the fit was converged), was fitted to Eq. 6 (Fig. 11(c)) within the temperature range of:

$$0.073 \leq \frac{T}{T_\omega} \leq 0.109 \tag{17}$$

A crucial point, why the fit still be able to converge, despite a narrow range of available $R(T)$ data (Eq. 17), is that the $R(T)$ dataset contained more than 12,000 raw resistive data points in the range indicated by Eq. 17. Thus, there is no surprize that the uncertainty in the deduced value of $T_\omega$ (Fig. 11(c)) is much larger than the uncertainty for $T_\theta$ (Fig. 11,b, $p = 5$), because the latter was deduced at nearly twice wider reduced temperature range:

$$0.138 \lesssim \frac{T}{T_\theta} \lesssim 0.205 \tag{18}$$

In addition, it should be mentioned that each additional free-fitting parameter in any fitting function causes an increase in the uncertainties of other free-fitting parameters.

Primary outcome of this part of the analysis is that $p$ and $T_\omega$ can be deduced for NRTS if experimental $R(T)$ datasets will be measured at much wider temperature range, which means that $R(T)$ dataset should not be limited by *de facto* the upper boundary of $T \sim 295$ K. In other words, $R(T)$ measurements should be performed at substantially wider $\frac{T}{T_\omega}$ range, and for the case of H$_3$S the upper limit should not be lower than $T_{upper\ exp} \sim 700\ K$ or $0.07 \lesssim \frac{T}{T_\omega} \lesssim 0.25$.

To prove this point of view, we can mention PdTe$_2$ (Section 3.3) and (ScZrNb)$_{0.65}$[RhPd]$_{0.35}$ (Section 3.4) compounds for which raw $R(T)$ data was measured within a range of $0.06 \lesssim \frac{T}{T_\omega} \lesssim 0.90$. In a result, both fits were converged and deduced $p$ and $T_\omega$ have small uncertainties (Figs. 3,4). From other hand, highly-compressed boron (Section 3.5) for which $R(T)$ data was measured at nearly twice narrower temperature range, $0.03 \lesssim$ $\frac{T}{T_\omega} \lesssim 0.55$, has larger uncertainties for $p$ and $T_\omega$ (Fig. 5) in comparison with PdTe$_2$ and (ScZrNb)$_{0.65}$[RhPd]$_{0.35}$.

Returning now to the sample 07/2018 we can report that *Im-3m*-phase of H$_3$S exhibits $T_\omega = 2707 \pm 184\ K$ and $p = 2.44 \pm 0.03$. The latter value is in close proximity to the electron-electron interaction value of $p = 2.0$.

This result can be explained as a manifestation of the covalent-like atomic bonds between ions in *Im-3m*-phase of H$_3$S. Detailed discussion of this feature of highly-pressurized sulfur hydride can be found elsewhere [58].

### 3.7. Highly-compressed LaH$_x$ (*P* = 150 GPa)

Drozdov *et al* [59] and Somayazulu *et al* [60] reported on the discovery of NRT superconductivity in highly-compressed lanthanum superhydride. NRTS phase was identified as *Fm-3m*-phase of La(H,D)$_{10}$. Due to the highest temperature at which *R*(*T*) data was measured for La-(H,D) compounds is $T_{upper\ exp} \sim 295\ K$ and a fact that La-(H,D) superconductors exhibit $T_c^{onset} \sim 240\ K$ (which is higher than $T_c^{onset} \sim 200\ K$ in H$_3$S) the problem of narrow $\frac{T}{T_\omega}$ range (discussed in Section 3.6) is more severe for La-(H,D) superconductors than for H$_3$S.

We fitted available *R*(*T*) datasets for La-(H,D) superconductors [58] to Eq. 5 when *p* = 5 in our earlier paper [34]. Here we report that there is only one *R*(*T*) dataset (from all available to date *R*(*T*) datasets [59]) of LaHx which can be fitted to Eq. 6 where *p* is a free-fitting parameter. This dataset is for Sample 11 [59]. This sample has hydrogen deficiency, LaH$_x$ (where x > 3) and exhibits *T*$_c$ ~ 70 K. The fits to Eqs. 6,8 are shown in Fig. 12.

It can be seen in Fig. 12(a), that the fit to Eq. 8 returns $N = 1.45 \pm 0.01$ which is approximately twice smaller than *N*-values deduced for H$_3$S samples (Figs. 6,8). At the same time, free-fitting parameter $p = 2.27 \pm 0.05$ is in close proximity to the value of $p = 2.44 \pm 0.03$ deduced for H$_3$S sample 07/2018 (Fig. 11,c). From this one can make a conclusion that Sample 11 of LaH$_x$ exhibits dominant electron-electron charge carrier interaction, similarly to highly-compressed boron and H$_3$S. We can point out that the electron-electron charge carrier interaction in LaH$_x$ is more likely also originated from the covalent-like atomic bonds.

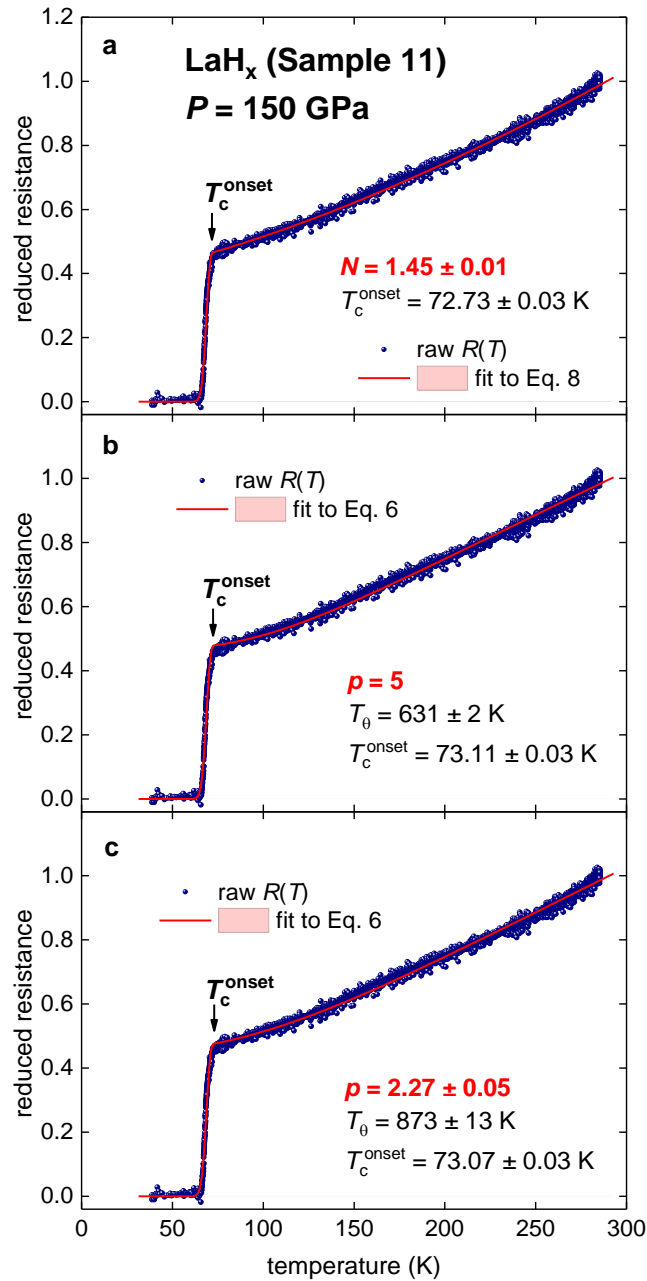

**Figure 12.** $R(T)$ data and fits to Eqs. 6,8 for highly-compressed hydrogen deficient LaH$_x$ (x > 3, Sample 11 [59]). (a) fit to Eq. 8, $N = 1.45 \pm 0.01$, 0.9979; (b) fit to Eq. 6, $p = 5$ (fixed), $T_\theta = 631 \pm 2\ K$, fit quality 0.9965; (c) fit to Eq. 6, $p = 2.27 \pm 0.05$, $T_\omega = 873 \pm 13\ K$, fit quality 0.9970. 95% confidence bands are narrower that the fitting curve.

It should be mentioned that deduced $p$ and $T_\omega$ have relative uncertainties within 2%. These high accuracies in deduced free-fitting parameters can be explained by:

1. $R(T)$ dataset was measured at a wide temperature range:

$$0.084 \lesssim \frac{T}{T_\omega} \lesssim 0.327 \tag{19}$$

where $T_{upper\ exp} = 285.45\ K$, $T_c^{onset} = 73.07 \pm 0.05\ K$, and $T_\omega = 873 \pm 13\ K$;

2. $R(T)$ dataset has more than 18,500 data points within the temperature range (Eq. 19).

### 3.8. Highly-compressed *F-43m*-PrH₉ and *P6₃/mmc*-PrH₉

Zhou *et al* [61] reported the discovery of low-$T_c$ superconductivity in highly-compressed praseodymium hydride. All synthesized samples contained a mixture of *F-43m*-PrH₉ and *P6₃/mmc*-PrH₉ phases.

Due to our results showed (Sections 3.1.-3.7) that power-law fit (Eqs. 7-8) is not reliable research tool, in Figure 13 we show only $R(T)$ fits to Eq.6 where $p$ is free-fitting parameter. measured at different pressure in the range of ($P$ = 135-145 GPa [61]).

In the studied pressure range ($P$ = 135-145 GPa [61]), the free-fitted parameter $p$ is varied in the range of $1.80 \leq p \leq 1.99$. This $p$ range is close to the theoretical value for pure electron-electron interaction, i.e. $p = 2.0$.

It should be mentioned that all three $R(T)$ datasets were measured at a wide temperature range:

$$0.01 \lesssim \frac{T}{T_\omega} \lesssim 0.25 \tag{20}$$

where $T_{upper\ exp} = 150\ K$, $T_c^{onset} \cong 7\ K$, and $T_\omega \cong 600\ K$, and $R(T)$, and for this reason, deduced $p$ and $T_\omega$ for all three samples have relative uncertainties less than 2%.

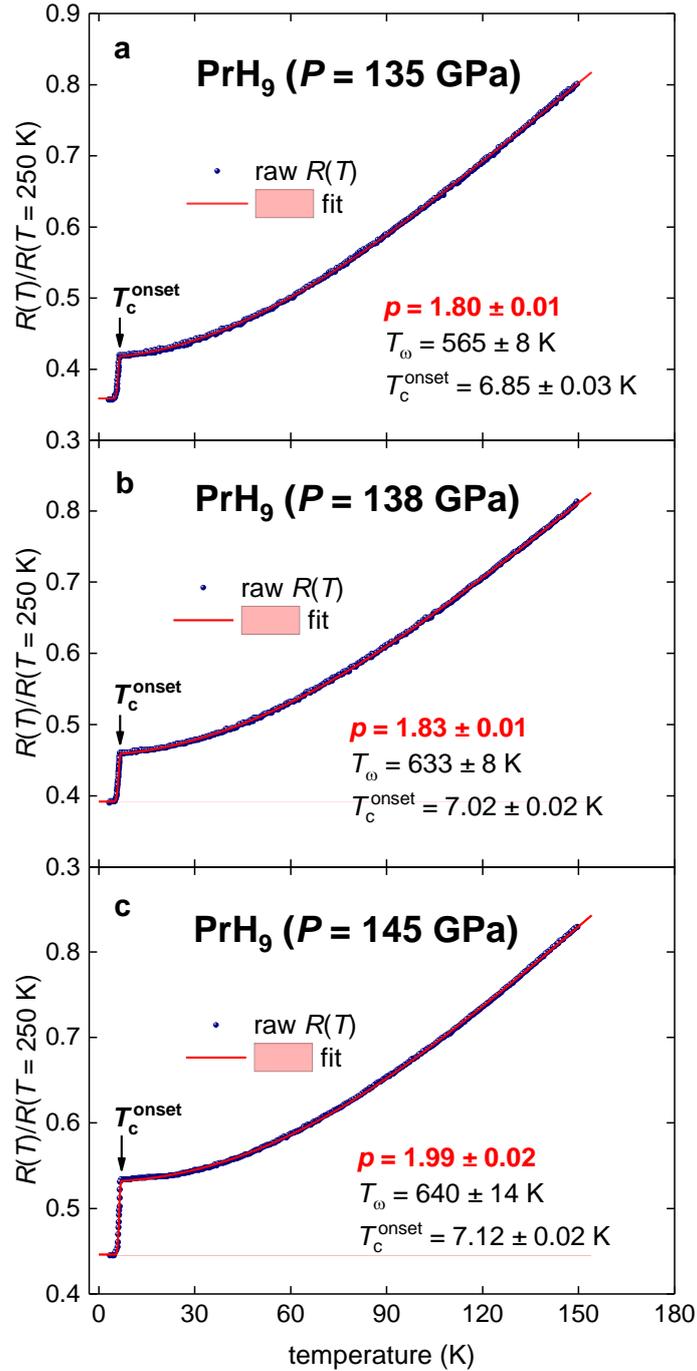

**Figure 13.** $R(T)/R(250\ K)$ data for highly-compressed $PrH_9$ ($P$ = 135-145 GPa [61]) and data fits to Eq. 6. (a) $P$ = 135 GPa, $p = 1.80 \pm 0.01$; fit quality 0.9998. (b) $P$ = 138 GPa, $p = 1.83 \pm 0.01$; fit quality 0.9998. (c) $P$ = 145 GPa, $p = 1.99 \pm 0.02$; fit quality 0.9997. 95% confidence bands are narrower that the fitting curve.

### 3.9. Highly-compressed pseudocubic $BaH_{12}$

Chen *et al* [62] reported the discovery of low-$T_c$ superconductivity in highly-compressed barium hydride $BaH_{12}$ which exhibits pseudocubic crystalline structure at pressure from 75 to 173 GPa. In Fig. 14 we analysed $R(T)$ curves (showed in Fig. S41(c) [45]) for the sample subjected to pressure $P = 100$ GPa and two consequent laser annealing/heating cycles. . It can be seen (Fig. 14,a) that $R(T)$ can be with a very good quality fitted for fixed $p = 5$ value. However, when $p$ is free-fitting parameter it decreased to the value close to pure case for the electron-magnon interaction, i.e. $p = 3$.

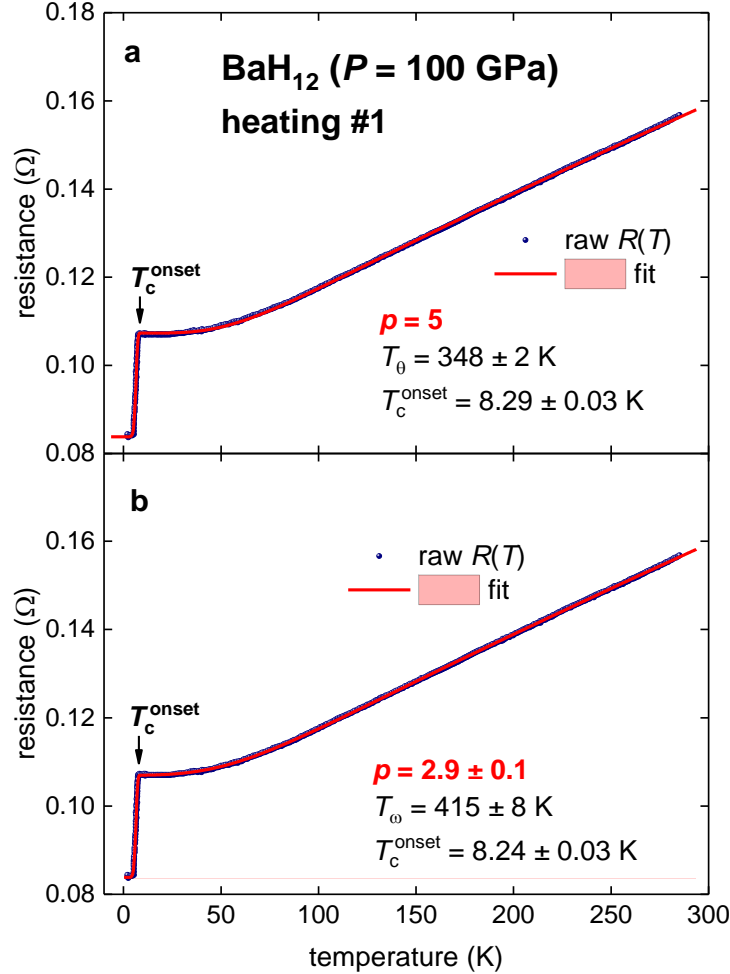

**Figure 14.** $R(T)$ data for highly-compressed $BaH_{12}$ (heating #1, $P = 100$ GPa [62]) and data fits to Eq. 6. (a) $p = 5$ (fixed); fit quality 0.9996; (b) $p = 2.9 \pm 0.1$; fit quality 0.9997.

Deduced $p$ and $T_\omega$ have relative uncertainties less than 3% (Fig. 14(b)) mainly because $R(T)$ dataset was measured at a wide temperature range:

$$0.02 \lesssim \frac{T}{T_\omega} \lesssim 0.67 \qquad (21)$$

where $T_{upper\ exp} = 280\ K$, $T_c^{onset} \cong 8\ K$, and $T_\omega = 415\ K$.

## 4. Discussion

Overall, in all considered highly-compressed superconductors, i.e. $\alpha$-Ga phase of boron, $H_3S$, $LaH_x$, $PrH_9$, $BaH_{12}$ and $\varepsilon$-phase of Fe (considered in Ref. 22), we have found essentially the same result that the power-law exponent, $p$, in the generalized Bloch-Grüneisen (BG) equation (Eq. 6) is well below 5, i.e. well below the value designated for pure electron-phonon charge carrier interaction in conductors.

On the other hand, for two conventional superconductors $ReBe_{22}$ [22] and high-entropy alloy $(ScZrNb)_{0.65}[RhPd]_{0.35}$ (Fig. 4) deduced free-fitting parameter $p$ is practically undistinguishable from 5. This implies that the electron-phonon interaction is the primary mechanism for emerge of superconductivity in $ReBe_{22}$ and $(ScZrNb)_{0.65}[RhPd]_{0.35}$. However, all highly-compressed superconductors exhibit different dominant charge carrier interaction and as a consequence of this non-electron-phonon pairing mechanism in the superconducting state.

It should be mentioned that there are many theoretically possible mechanisms for charge carrier pairing in the condensed matter [11-14], and from the deduced power-law exponent presented herein, it is more likely that the primary mechanism is related to the electron-electron interaction. This finding is in good accord with the empirical finding that all hydrogen-rich superconductors are located [17,18,46,63] in the unconventional superconductors region of the Uemura plot [19,20].

Taking into account a possible argument that first-principles calculations have predicted reasonably accurately the superconducting transition temperatures of several (and what is important to stress, not of all) hydrogen-rich superconductors, the discussion can be related to the issue that sufficiently strong electron-phonon interaction can be accommodated as consequence of primary and very strong electron-electron interaction. Thus, by varying the Coulomb pseudopotential parameter, μ*, and the electron-phonon coupling constant, $\lambda_{e-ph}$, in equations similar to the ones proposed by McMillan [54], and Allen and Dynes [55-57]:

$$T_c = \left(\frac{1}{1.20}\right) \cdot \left(\frac{\hbar}{k_B}\right) \cdot \omega_{ln} \cdot e^{-\left(\frac{1.04 \cdot (1+\lambda_{e-ph})}{\lambda_{e-ph} - \mu^* \cdot (1+0.62 \cdot \lambda_{e-ph})}\right)} \cdot f_1 \cdot f_2 \qquad (6)$$

where $k_B$ is the Boltzmann constant, and $\hbar$ is the reduced Planck constant, $\omega_{ln}$ is logarithmic phonon frequency, and $f_1$ and $f_2$ are correction functions, it is possible to find correlative dependence between a primary mechanism (which is non-electron-phonon) with the associated electron-phonon interaction.

There is a necessity to discuss the applicability of the high-temperature resistance saturation model [64,65] to the NRST superconductors. The model [64,65] has been applied to describe the resistivity of many metallic alloys when the electron mean-free path, $l$, is of the order of the interatomic spacing, $a$, in the compound. It was shown in Ref. 63, that H$_3$S exhibits the inequality:

$$a \ll \xi(0) = 2.5 nm \ll l(T \sim 200K). \qquad (23)$$

Thus, the high-temperature resistance saturation model [64,65] cannot be applicable for the analysis of H$_3$S samples because the condition of $a \cong l$ does not satisfy.

Speaking more broadly, in metallic compounds the condition of $a \cong l$ is satisfied when $T > T_\theta$. If one takes into account that available to date $R(T)$ datasets for NRST is limited by $\frac{T_{upper\ exp}}{T_\theta} \lesssim 0.205$ (Eq. 18) and a fact that diamond anvil cells cannot operate at temperatures of $T > T_\theta \geq 1,400\ K$, it is clear that the high-temperature resistance saturation model [64,65] is not applicable for the analysis of $R(T)$ curves of NRST superconductors. However, there is a possibility that the model [64,65] can be applicable for some low-$T_c$ hydrogen-rich superconductors (for instance, BaH$_{12}$ for which deduced $T_\theta = 348\ K$ (Section 3.9)) if experimental $R(T)$ curves will be measured.

There is another important issue that has been never discussed in the literature. This is a possibility, that NRTS phases are amorphous. Taking in account that NRTS superconductors are synthesised at extreme heating/colling rates at high pressure, there is a chance that the NRTS state exhibits in amorphous phases which co-exist with the crystalline phases. The crystalline hydrogen-rich phases can be detected by XRD, however, there is a challenging task to detect amorphous hydrogen-rich phases in samples inside of diamond anvil cell.

There is a well-established experimental fact that amorphous tungsten exhibits superconducting transition temperature of $T_c = 4.8$ K [66], while its crystalline counterpart has $T_c = 0.012$ K [67]. In this regard it is important to mention recently discovered high-pressure $Ge_2Sb_2Te_5$ superconductor which exhibits the superconducting state in amorphous phase at nearly ambient pressure in the decompression run [68].

This means that the translational symmetry, the key assumption of the BCS and Eliashberg's theories, does not necessary condition for the superconductivity as physical phenomenon. Recent experimental discovery of the superconductivity in quasicrystals [69], where the translational symmetry does not exist supports this point of view.

## 5. Conclusions

In conclusion, in this paper we analysed $R(T)$ data for layered transition metal dichalcogenides $PdTe_2$, high-entropy alloy $(ScZrNb)_{0.65}[RhPd]_{0.35}$, and highly-compressed α-Ga phase of boron, $H_3S$, $LaH_x$, $PrH_9$ and $BaH_{12}$ and showed that the dominant charge carrier interaction in studied highly-compressed superconductors has non-electron-phonon nature.

**Acknowledgement**

The author thanks Dr. K. Stolze (Leibniz-Institut für Kristallzüchtung) for providing $R(T)$ data for high-entropy alloy $(ScZrNb)_{0.65}[RhPd]_{0.35}$, Dr. M. I. Eremets and Dr. V. S. Minkov (Max-Planck Institut für Chemie, Mainz, Germany) for providing $R(T)$ data for *Im-3m*-phase of sulphur hydride and for *Fm-3m*-phase of lanthanum hydride, and Dr. S. Mozafarri and co-authors (National High Magnetic Field Laboratory, Florida State University, USA) for open access magnetoresistance data for *Im-3m*-phase of $H_3S$ [34]. The author thanks one anonymous Referee for proofreading of the manuscript.

The author thanks financial support provided by the Ministry of Science and Higher Education of Russia (theme "Pressure" No. AAAA-A18-118020190104-3) and by Act 211 Government of the Russian Federation, contract No. 02.A03.21.0006.

## References

[1]  Satterthwaite C B and Toepke I L 1970 Superconductivity of hydrides and deuterides of thorium *Phys. Rev. Lett.* **25** 741-743
[2]  Ashcroft N W 1968 Metallic hydrogen: a high-temperature superconductor? *Phys. Rev. Lett.* **21** 1748-1749
[3]  Bardeen J, Cooper L N, and Schrieffer J R 1957 Theory of superconductivity *Phys. Rev.* **108** 1175-1204
[4]  Eliashberg G M 1960 Interactions between electrons and lattice vibrations in a superconductor *Soviet Phys. JETP* **11** 696-702
[5]  Ashcroft N W 2004 Hydrogen dominant metallic alloys: high temperature superconductors? *Phys. Rev. Lett.* **92** 187002
[6]  Stritzker B and Buckel W 1972 Superconductivity in the palladium-hydrogen and the palladium-deuterium systems *Zeitschrift für Physik A Hadrons and nuclei* **257** 1-8
[7]  Yussouff M, Rao B K and Jena P 1995 Reverse isotope effect on the superconductivity of PdH, PdD, and PdT *Solid State Communications* **94** 549-553
[8]  Drozdov A P, Eremets M I, Troyan I A, Ksenofontov V, Shylin S I 2015 Conventional superconductivity at 203 kelvin at high pressures in the sulfur hydride system *Nature* **525** 73-76
[9]  Minkov V S, Prakapenka V B, Greenberg E and Eremets M I 2020 Boosted $T_c$ of 166 K in superconducting $D_3S$ synthesized from elemental sulfur and hydrogen *Angew. Chem. Int. Ed.* **59** 18970-18974
[10]  Drozdov A P, *et al* 2019 Superconductivity at 250 K in lanthanum hydride under high pressures *Nature* **569** 528-531
[11]  Monthoux P, Pines D and Lonzarich G G 2007 Superconductivity without phonons *Nature* **450** 1177-1183
[12]  Sachdev S 2012 What can gauge-gravity duality teach us about condensed matter physics? *Annual Review of Condensed Matter Physics* **3** 9-33
[13]  Matthias B T 1971 Anticorrelations in superconductivity *Physica* **55** 69-72
[14]  [38]  Kim H-T 2021 Room-temperature-superconducting $T_c$ driven by electron correlation *Scientific Reports* **11** 10329

[15] Bloch F 1930 Zum elektrischen Widerstandsgesetz bei tiefen Temperaturen *Z. Phys.* **59** 208-214
[16] Grüneisen E 1933 Die abhängigkeit des elektrischen widerstandes reiner metalle von der temperatur. *Ann. Phys.* **408** 530–540
[17] Talantsev E F 2019 Classifying superconductivity in compressed $H_3S$ *Modern Physics Letters B* **33** 1950195
[18] Talantsev E F 2020 An approach to identifying unconventional superconductivity in highly-compressed superconductors *Superconductor Science and Technology* **33** 124001
[19] Uemura Y J 1997 Bose-Einstein to BCS crossover picture for high-$T_c$ cuprates *Physica C* **282-287** 194-197
[20] Uemura Y J 2019 Dynamic superconductivity responses in photoexcited optical conductivity and Nernst effect *Phys. Rev. Materials* **3** 104801
[21] Poker D B and Klabunde C E 1982 Temperature dependence of electrical resistivity of vanadium, platinum, and copper *Phys. Rev. B* **26** 7012
[22] Talantsev E 2021 Quantifying the charge carrier interaction in metallic twisted graphene superlattices *Nanomaterials* **11** 1306
[23] Jiang H, *et al* 2015 Physical properties and electronic structure of $Sr_2Cr_3As_2O_2$ containing $CrO_2$ and $Cr_2As_2$ square-planar lattices *Phys. Rev. B* **92** 205107
[24] Shang T, et al 2019 Enhanced $T_c$ and multiband superconductivity in the fully-gapped $ReBe_{22}$ superconductor *New J. Phys.* **21** 073034
[25] Barker J A T, *et al* 2018 Superconducting and normal-state properties of the noncentrosymmetric superconductor $Re_3Ta$ *Phys. Rev. B* **98** 104506
[26] Matsumoto R, *et al* 2019 Pressure-induced superconductivity in tin sulphide *Phys. Rev. B* **99** 184502
[27] Matula R A 1979 Electrical resistivity of copper, gold, palladium, and silver *J. Phys. Chem. Ref. Data* **8** 1147-1298
[28] Teixeira J 1974 Resistivity of pure metals and dilute magnetic alloys *Ph.D. Thesis* (University of Science and Medicine and National Polytechnical University of Grenoble, Saint-Martin-d'Hères, France) 1-155
[29] White G K and Woods S B 1959 Electrical and thermal resistivity of the transition elements at low temperatures *Phil. Trans. R. Soc. Lond. A* **251**, 273-302
[30] Shimizu K, et al 2001 Superconductivity in the nonmagnetic state of iron under pressure *Nature* **412** 316-318
[31] Jaccard D, *et al* 2002 Superconductivity of ε-Fe: complete resistive transition *Physics Letters A* **299** 282-286
[32] Polshyn H, *et al* 2019 Large linear-in-temperature resistivity in twisted bilayer graphene *Nature Physics* **15** 1011
[33] Talantsev E F and Stolze K 2021 Resistive transition in hydrogen-rich superconductors *Superconductor Science and Technology* **34** 064001
[34] Talantsev E F 2020 Advanced McMillan's equation and its application for the analysis of highly-compressed superconductors *Superconductor Science and Technology* **33** 094009
[35] Ohta K, Kuwayama Y, Hirose K, Shimizu K, Ohishi Y. 2016 Experimental determination of the electrical resistivity of iron at Earth's core conditions *Nature* **534** 95-98
[36] Gschneidner K A Jr 1964 Physical properties and interrelationships of metallic and semimetallic elements *Solid State Physics* **16** 275-426
[37] Sipos B, et al. 2008 From Mott state to superconductivity in $1T-TaS_2$. *Nat. Mater.* **7** 960-965
[38] Park S, et al 2021 Superconductivity emerging from a stripe order in $IrTe_2$ nanoflakes *Nat. Comms.* **12** 3157

[39] Wilson J A, Salvo F J D and Mahajan S 1975 Charge-density waves and superlattices in the metallic layered transition metal dichalcogenides *Adv. Phys.* **24** 117-201

[40] Yang H, *et al* 2021 Anomalous charge transport of superconducting $Cu_xPdTe_2$ under high pressure *arXiv*: 2106.05613

[41] Stolze K, Tao J, von Rohr F O, Kong T, Cava R J 2018 Sc−Zr−Nb−Rh−Pd and Sc−Zr−Nb−Ta−Rh−Pd high-entropy alloy superconductors on a CsCl-type lattice *Chem. Mater.* **30** 906-914

[42] Eremets M I, Struzhkin V V, Mao H-K, Hemley R J 2001 Superconductivity in boron *Science* **293** 272-274

[43] Oganov A R, *et al* 2009 Ionic high-pressure form of elemental boron *Nature* **457** 863-867

[44] Mozaffari S, *et al* 2019 Superconducting phase diagram of $H_3S$ under high magnetic fields *Nature Communications* **10** 2522

[45] Capitani F, et al. 2017 Spectroscopic evidence of a new energy scale for superconductivity in $H_3S$ *Nat. Phys.* **13** 859-863

[46] Talantsev E F 2021 Comparison of highly-compressed $C2/m$-$SnH_{12}$ superhydride with conventional superconductors *Journal of Physics: Condensed Matter* **33** 285601

[47] Li Y, *et al* 2014 The metallization and superconductivity of dense hydrogen sulfide *The Journal of Chemical Physics* **140** 174712

[48] Duan D, *et al* 2015 Pressure-induced metallization of dense $(H_2S)_2H_2$ with high-$T_c$ superconductivity *Scientific Reports* **4** 6968

[49] Duan D, *et al* 2015 Pressure-induced decomposition of solid hydrogen sulphide *Phys. Rev. B* **91** 180502

[50] Errea I *et al* 2015 High-pressure hydrogen sulfide from first principles: A strongly anharmonic phonon-mediated superconductor *Phys. Rev. Lett.* **114** 157004

[51] Durajski A P 2016 Quantitative analysis of nonadiabatic effects in dense $H_3S$ and $PH_3$ superconductors *Sci. Rep.* **6** 38570

[52] Goncharov A F, Lobanov S S, Kruglov I, Zhao X-M, Chen X-J, Oganov A R, Konopkova Z, Prakapenka V B 2016 Hydrogen sulfide at high pressure: Change in stoichiometry *Phys. Rev. B* **93** 174105

[53] Zureka E and Bi T 2019 High-temperature superconductivity in alkaline and rare earth polyhydrides at high pressure: A theoretical perspective *J. Chem. Phys.* **150** 050901

[54] McMillan W L 1968 Transition temperature of strong-coupled superconductors *Phys. Rev.* **167** 331-344

[55] Dynes R C 1972 McMillan's equation and the $T_c$ of superconductors *Solid State Communications* **10** 615-618

[56] Allen P B and Dynes R C 1975 Transition temperature of strong-coupled superconductors reanalysed *Phys. Rev. B* **12** 905-922

[57] Allen P B, in *Handbook of Superconductivity* (edited by C. P. Poole, Jr.) (Academic Press, New York, 1999) Ch. 9, Sec. G, pp. 478-489

[58] Gor'kov L P and Kresin V Z 2018 Colloquium: High pressure and road to room temperature superconductivity *Review of Modern Physics* **90** 011001

[59] Drozdov A P, *et al* 2019 Superconductivity at 250 K in lanthanum hydride under high pressures *Nature* **569** 528-531

[60] Somayazulu M, *et al* 2019 Evidence for superconductivity above 260 K in lanthanum superhydride at megabar pressures *Phys. Rev. Lett.* **122** 027001

[61] Zhou D, *et al* 2020 Superconducting praseodymium superhydrides *Science Advances* **6** eaax6849

[62] Chen W, *et al* 2021 Synthesis of molecular metallic barium superhydride: pseudocubic $BaH_{12}$ *Nature Communications* **12** 273

[63] Talantsev E F 2019 Classifying hydrogen-rich superconductors *Materials Research Express* **6** 106002

[64] Fisk Z and Webb G W 1976 Saturation of the high-temperature normal-state electrical resistivity of superconductors *Phys Rev Lett* **36** 1084-1086

[65] Wiesmann H, *et al* 1977 Simple model for characterizing the electrical resistivity in A-15 superconductors *Phys Rev Lett* **38** 782-785

[66] Sun Y, *et al* 2013 Voltage-current properties of superconducting amorphous tungsten nanostrips *Scientific Reports* **3** 2307

[67] Gibson J W and Hein R A 1964 Superconductivity of tungsten *Phys. Rev. Lett.* **12** 688

[68] Huang G, *et al* 2021 Memory of pressure-induced superconductivity in a phase-change alloy *Phys Rev B* **103** 174515

[69] Kamiya K, Takeuchi T, Kabeya N, Wada N, Ishimasa T, Ochiai A, Deguchi K, Imura K and Sato N K 2018 Discovery of superconductivity in quasicrystal *Nature Communications* **9** 154